\documentclass{article}              
\usepackage{graphicx}
\usepackage{xcolor}
\usepackage{amsmath}
\usepackage{authblk}

\begin{document}                  



\title{Using low dose X-ray Speckle Visibility Spectroscopy to study dynamics of soft matter samples}


\author[a]{Johannes M\"oller}
\author[a,b]{Mario Reiser}
\author[a]{J\"org Hallmann}
\author[a]{Ulrike Boesenberg}
\author[a]{Alexey Zozulya}
\author[b]{Hendrik Rahmann}
\author[b]{Anna-Lena Becker}
\author[c]{Fabian Westermeier}
\author[d]{Thomas Zinn}
\author[c]{Michael Sprung}
\author[d]{Theyencheri Narayanan}
\author[b]{Christian Gutt}
\author[a]{Anders Madsen}

\affil[a]{European X-ray Free Electron Laser Facility, Holzkoppel 4, 22869 Schenefeld, Germany}
\affil[b]{University Siegen, 57076 Siegen, Germany}
\affil[c]{Deutsches Elektronen Synchrotron DESY, 22607 Hamburg,  Germany}
\affil[d]{ESRF - The European Synchrotron, 38000 Grenoble,  France}



\maketitle                        

\begin{abstract}
We demonstrate the successful application of X-ray Speckle Visibility Spectroscopy (XSVS) experiments to study the dynamics of radiation sensitive, biological samples with unprecedentedly small X-ray doses of 45 Gy and below. Using XSVS, we track the dynamics of casein micelles in native, concentrated, and acidified solution conditions, while substantially reducing the deposited dose as compared to alternative techniques like sequential X-ray photon correlation spectroscopy (XPCS). The Brownian motion in a skim milk sample yields the hydrodynamic radius of the casein micelles while deviations from Brownian motion with a characteristic $q$-dependent diffusion coefficient $D(q)$ can be observed in more concentrated solution conditions. The low dose applied in our experiments allows the observation of static, frozen speckle patterns from gelled acidic milk. We show that the XSVS technique is especially suitable for tracking dynamics of radiation sensitive samples in combination with the improved coherent properties of new generation X-ray sources, emphasizing the great potential for further investigations of protein dynamics using fourth generation synchrotrons and free electron lasers.

\end{abstract}


\section{Introduction}


X-ray photon correlation spectroscopy (XPCS) provides access to a hierarchy of spatial and temporal information with the exciting outlook of extending the application of biological Small-Angle X-ray Scattering (Bio-SAXS) experiments considerably \cite{Vodnala2018,Moeller2019a,Begam2021,girelli_microscopic_2021} by adding dynamical information. As a coherent scattering technique, XPCS will especially benefit from highly coherent X-ray sources such as 4th generation synchrotron storage rings and MHz X-ray free-electron lasers \cite{Einfeld2014,Weckert2015,Raimondi2016,Tschentscher2017,Schroer2018,Decking2020}. Thus, Bio-XPCS methods bear the potential of time-resolved investigations of processes relevant to biological function such as aggregation, phase separation, partitioning, and self-assembly, which are important for understanding for instance neurodegenerative diseases, the development of tissues and organs, virus infections, the formation of bio-inspired materials and many more. Moreover, studying dynamics in crowded biological systems can provide fundamental insight into protein interactions in concentrated environments \cite{Stradner2020,grimaldo_2019} advancing our microscopic understanding of processes such as misfolding and aggregation related to pathological cases such as Alzheimer’s and prion diseases. It can also help in the design and implementation of improved protein-based therapeutics with tailored transport and storage properties.

While the science case of Bio-XPCS is compelling, the technical challenges of realizing such experiments are considerable with radiation damage being one of the main obstacles when using highly intense X-ray beams. Atomic scale XPCS experiments, for example, apply X-ray doses of up to MGy and beyond, which can lead to beam-induced dynamics even in hard-condensed-matter samples \cite{Ruta2017,Pintori2019,Holzweber2019}. Soft and biological matter samples are much more sensitive to radiation damage requiring continuous flowing \cite{Fluerasu2008,westermeier2012,Vodnala2018} or scanning of the samples \cite{Lurio2021} with optimized data-taking strategies \cite{Verwohlt2018}. Typical critical X-ray doses for protein molecules in solution range from 7 - 10 kGy (BSA) to 0.3 kGy (RNase) after which a degradation of the SAXS patterns is visible \cite{Jeffries2015}. These doses are easily reached within milliseconds when using focused beams of modern synchrotron sources. 

In order to overcome this bottleneck of radiation damage, adapted experimental XPCS schemes such as serial X-ray Speckle Visibility Spectroscopy (XSVS) \cite{Gutt2009,Hruszkewycz2012,Inoue2012,Li2014} are needed, which allow to make full use of the increased brilliance and coherence of the new X-ray sources. An additional advantage of XSVS is that the temporal resolution is given by the single frame exposure time and not by the frame rate as in XPCS, therefore having less stringent requirements to the readout speed of pixelated X-ray detectors.

Here, we demonstrate the feasibility of measuring dynamics in biological systems with ultra low dose XSVS experiments. We make use of short exposure times and measure the dynamics of casein micelles with a maximum applied X-ray dose of 0.045 kGy only. This dose is orders of magnitude below damage thresholds reported for many monomeric proteins. Additionally, the results allow us to estimate the feasibility of Bio-XSVS experiments for a larger variety of proteins on a wide range of length scales.

\section{Low dose coherent X-ray scattering experiments}

The experiments were performed at the coherent applications beamline P10, PETRA III, DESY, Hamburg and at the time-resolved (U-)SAXS beamline ID02, ESRF, France \cite{Zinn2018}. The casein micelle samples were derived from commercial skim milk powder (SUCOFIN), and used without further purification. Residual aggregates and lipids were removed by centrifugation. A "native" sample was prepared following the supplier's instruction, resulting in a final casein protein concentration of about 3 wt$\%$.

The scattered intensity was recorded with two different versions of the EIGER photon counting detector \cite{Dinapoli2011,Radicci2012,Johnson2014}. Structural information can be obtained by azimuthally integrating the scattered intensity on the detector. The obtained scattering intensity $I(q)$, with $q = 4 \pi \sin (\theta) / \lambda$ designating the modulus of the scattering vector, the wavelength $\lambda$ and the scattering angle $2 \theta$, contains information about the size, shape and size distribution of the scattering entities in the sample. The shape of the scattering particles is given by the single particle form factor $P(q)$, which in the simple case of spheres with a homogeneous scattering density and radius $R$ is
\begin{equation}
    P(q) =  \left( 3\frac{\sin(qR)-qR \cdot \cos(qR)}{(qR)^3} \right)^2. \label{eq:form}
\end{equation}

An additional modulation (speckle) of the scattered intensity can be observed for coherent scattering experiments, when the coherence volume of the incident radiation becomes comparable to the scattering volume. This speckle pattern depends on the spatial arrangement of the scatterers and therefore dynamical information can be obtained from the temporal evolution of the speckle pattern.

The size of the speckles on the detector can be estimated as $S = \frac{\lambda L}{a}$
with $a$ being the X-ray spot size on the sample and $L$ the sample-detector-distance. To minimize the absorbed dose, the measurements were performed with comparably large $a$, which consequently requires large $L$ to resolve the speckles on the detector. A more detailed discussion about the best compromise between speckle contrast and dose can be found in \cite{Moeller2019a}. A summary of the experimental parameters used in this study is given in table \ref{tbl:2}.

\begin{table}
  \caption{Experimental setups for low dose coherent X-ray scattering experiments.}
  \label{tbl:2}
  \begin{tabular}{l|l|l}
    \hline
    beam-line & P10, DESY
 & ID02, ESRF \\
     \hline
     detector & Dectris Eiger 4M
 & PSI Eiger 500k
 \\ frame rate & 750 Hz
 & 22 kHz
 \\ L & 20 m
 & 30.7 m
 \\ wavelength & 1.54 \AA
 & 0.995 \AA
 \\ beamsize (a x a) & $\approx$ 60 $\mu$m x 60 $\mu$m
 & $\approx$ 30 $\mu$m x 30 $\mu$m \\ flux & $\approx$ $2 \ 10^{10}$ ph/s
 & $\approx$ $4 \ 10^{10}$ ph/s \\ $\beta_0$ & $\approx$  23 \%
 & $\approx$ 28 \% \\ 
 $q$-range & $2.5 \ 10^{-3} \ \mathrm{nm}^{-1}$
 & $2 \ 10^{-3} \ \mathrm{nm}^{-1}$ \\
   & $ - \ 2 \ 10^{-1} \ \mathrm{nm}^{-1}$
 & $ - 1.5 \ 10^{-1} \ \mathrm{nm}^{-1}$ \\
    \hline
  \end{tabular}
\end{table}

\subsection{Speckle visibility spectroscopy in the low count regime}

In standard XPCS experiments, sequential acquisitions from the same sample spot are measured, so that the intermediate scattering function $g_{1}(q,t^{\prime})$ (ISF) can be calculated from the scattered intensity $I(\vec{q})$ as 
\begin{equation}
g_2 (q,t^{\prime}) = \frac{\left\langle I(\vec{q},t)I(\vec{q},t+t^{\prime})\right\rangle}{\left\langle I(\vec{q})\right\rangle^2} = 1 + \beta_0 \left|g_{1}(q,t^{\prime})\right|^2,\label{eq:g2}
\end{equation}
with $\beta_0$ being the speckle contrast. The time delay between two consecutive time frames is denoted $t^{\prime}$ and $\left\langle \ldots \right\rangle$ is the ensemble average over all equivalent delay times $t$ and pixels within a certain range of the absolute value $\left|\vec{q}\right|$.\\
In order to reduce the deposited dose on a single spot of the sample as much as possible, we aim at obtaining the same dynamic information from single acquisitions. As the speckle contrast $\beta$ depends on the exposure time $\tau$ and the ISF of the sample like \cite{Bandy2005}
\begin{equation}
\beta(q,\tau) = 2 \frac{\beta_0}{\tau} \int_0^\tau (1-t^{\prime}/\tau) \left|g_{1}(q,t^{\prime})\right|^2 \mathrm{d}t^{\prime},\label{eq:bandy}
\end{equation}
one can probe the properties of the ISF from a series of single frame acquisitions by varying the exposure time.\\
The speckle contrast for each acquisition can be calculated from a statistical analysis of the photon counts. Following the negative-binomial distribution \cite{Goodman1985}, also known as the Poisson-Gamma distribution, the speckle contrast can be obtained from the probabilities $P(k)$ of pixels within an ensemble measuring $k=0, 1, 2 \ldots$ photons as
\begin{equation}
	\beta = \frac{P(0)}{P(1)} - \frac{1}{\left\langle k\right\rangle},\label{eq:beta}
\end{equation}
with $\left\langle k\right\rangle$ denoting the mean number of photons per pixel per acquisition. The error in determining the contrast is given by \cite{Verwohlt2018}
\begin{equation}
	\Delta \beta = \frac{1}{\left\langle k\right\rangle} \sqrt{\frac{2(1+\beta)}{n_{pix} n_{im}}},\label{eq:beta_err}
\end{equation}
with $n_{pix}$ the number of pixel in the region of interest and $n_{im}$ the number of frames. Determining the speckle contrast with high accuracy is the main difficulty of XSVS experiments with very low count rates. Therefore, extensive characterization of the experimental setup was performed before the experiments using a static sample. Substantial deviations in the measured photon statistics from those of the expected Poisson-Gamma distribution at low count rates were found for both types of EIGER detectors used for this experiment. Due to this non-ideal behavior, we developed a detector correction scheme, which is explained in detail in \cite{Moeller2019b}. Performing this correction is mandatory for the experiments to work and was applied to all measurements in the following. \\

\begin{figure}
  \includegraphics[width=0.95\textwidth]{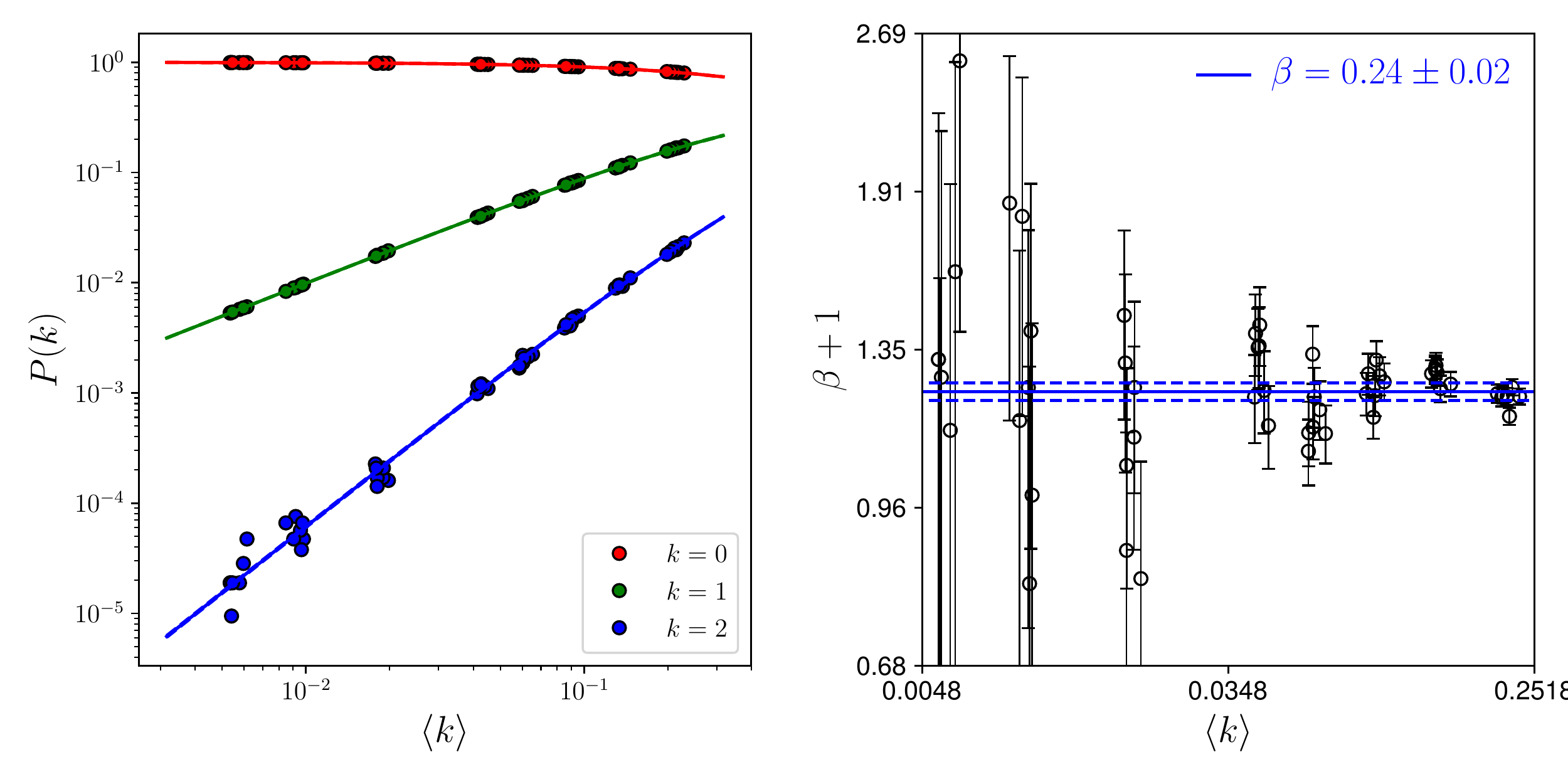}
  \caption{(left) Probabilities $P(k)$ of measuring k = 0, 1 or 2 photons at a fixed exposure time of $\tau = 1.35$ $\mu$s and at a scattering vector $q = (6 \pm 2) \ 10^{-3} \ \mathrm{nm}^{-1}$, which translates into a region of 1230 pixels on the detector (at beamline P10 with 20 m sample-to-detector distance, native casein solution). The incident intensity was varied between $2 \ 10^{8}$ ph/s and $2 \ 10^{10}$ ph/s and the data was corrected following \cite{Moeller2019b}. Each data point is calculated as the mean of $86$ single acquisitions with closely matching $\left\langle k\right\rangle$. The solid lines represent the Poisson-Gamma distribution, with a contrast $\beta = 0.24$. (right) Resulting contrast values, calculated following eq. \ref{eq:beta}. The contrast value $\beta$ is obtained as a weighted average and displayed as a solid line. The uncertainty of the fit is displayed as a dashed line band.}
  \label{fgr:beta}
\end{figure}
An exemplary data set is displayed in figure \ref{fgr:beta}, where the incident intensity was varied by using silicon attenuators, in order to investigate the obtainable SNR as a function of coherent flux. On the left hand side, the probability $P(k)$ of measuring $k$ = 0, 1 or 2 photons is plotted as a function of mean scattered intensity per pixel $\left\langle k\right\rangle$. The resulting speckle contrast is calculated for each triplet of $P(0)$, $P(1)$ and $\left\langle k\right\rangle$ values, following eq. \ref{eq:beta}, with the error following eq. \ref{eq:beta_err} (fig. \ref{fgr:beta}, right). The final $\beta(q,\tau)$ is calculated as a weighted average of the single data points and displayed as solid blue line, with dashed lines representing the statistical uncertainties stemming from the weighted average calculation. It can be clearly seen that with increasing mean intensity $\left\langle k\right\rangle$ the statistical error of $\beta$ reduces significantly. Therefore, the number of acquired images ($n_{im}$) was adapted in the following for each exposure time according to the obtained mean intensity $\left\langle k\right\rangle$. Additionally, the beam attenuation was adapted, in order to avoid exceeding a deposited dose of $45$ Gy per acquisition. As can be seen from eq. \ref{eq:beta_err}, in order to preserve the statistical accuracy of the speckle contrast, a reduction of the intensity by a factor 10 requires an increase in the number of images by a factor of 100. A full list of used exposure times and corresponding doses can be found in Tab. \ref{tbl:1}.\\
 
 \begin{table}
  \caption{Number of images $n_{im}$ and dose per acquisition $\tilde{D}$ measured at beamline ID02.}
  \label{tbl:1}
  \begin{tabular}{l|llllllll}
    \hline
     exposure time $\tau$ [ms] & 0.1 & 0.3 & 1 & 3 & 10 & 30 & 100 & 300\\
     \hline
     $n_{im}$ & 21,600 & 18,000 & 12,000 & 6,000 & 2,160 & 3,840 & 1,080 & 360  \\    
     attenuator transmission & 1 & 1 & 1 & 1 & 1 & 0.05 & 0.05 & 0.05 \\
     $\tilde{D}$ [Gy] & 0.3 & 0.9 & 3 & 9 & 30 & 4.5 & 15 & 45  \\ 
     $\left\langle k\right\rangle ( q = 0.8 \ 10^{-2} \mathrm{nm}^{-1} )$ & 0.030 & 0.089 & 0.29 & 0.89 & 2.95 & 0.44 & 1.47 & 4.43  \\ 
     $\left\langle k\right\rangle ( q = 2.2 \ 10^{-2} \mathrm{nm}^{-1} )$ & 0.007 & 0.021 & 0.07 & 0.21 & 0.70 & 0.11 & 0.35 & 1.05  \\ 
    \hline
  \end{tabular}
\end{table}

\section{Dynamics of casein micelles obtained by low dose XSVS}

\subsection{Dynamics of native Casein micelle solutions}

We first investigate a native casein micelle solution obtained from commercially available skim milk powder. The structure of casein micelles has been investigated extensively with X-ray scattering methods, using different approaches for modeling and interpretation of the data due to the hierarchical structure \cite{horne2006,Shukla2009,deKruif2012,Ingham2016}.
Dynamics of casein micelles have primarily been studied by light scattering methods and rheology \cite{mezzenga2005}. Due to the strong turbidity of the sample, these measurements were either limited to diluted samples, used advanced cross-correlation schemes to suppress the strong multiple scattering contribution \cite{urban1999}, or where restricted to the multiple scattering regime using diffusing wave spectroscopy (DWS) \cite{dahbi2010,alexander2008}. In this context, XSVS in the ultra-small angle regime offers the advanced capability of obtaining the full $q$- and direction-dependent information of such concentrated soft matter samples.

\begin{figure}
  \includegraphics[width=0.95\textwidth]{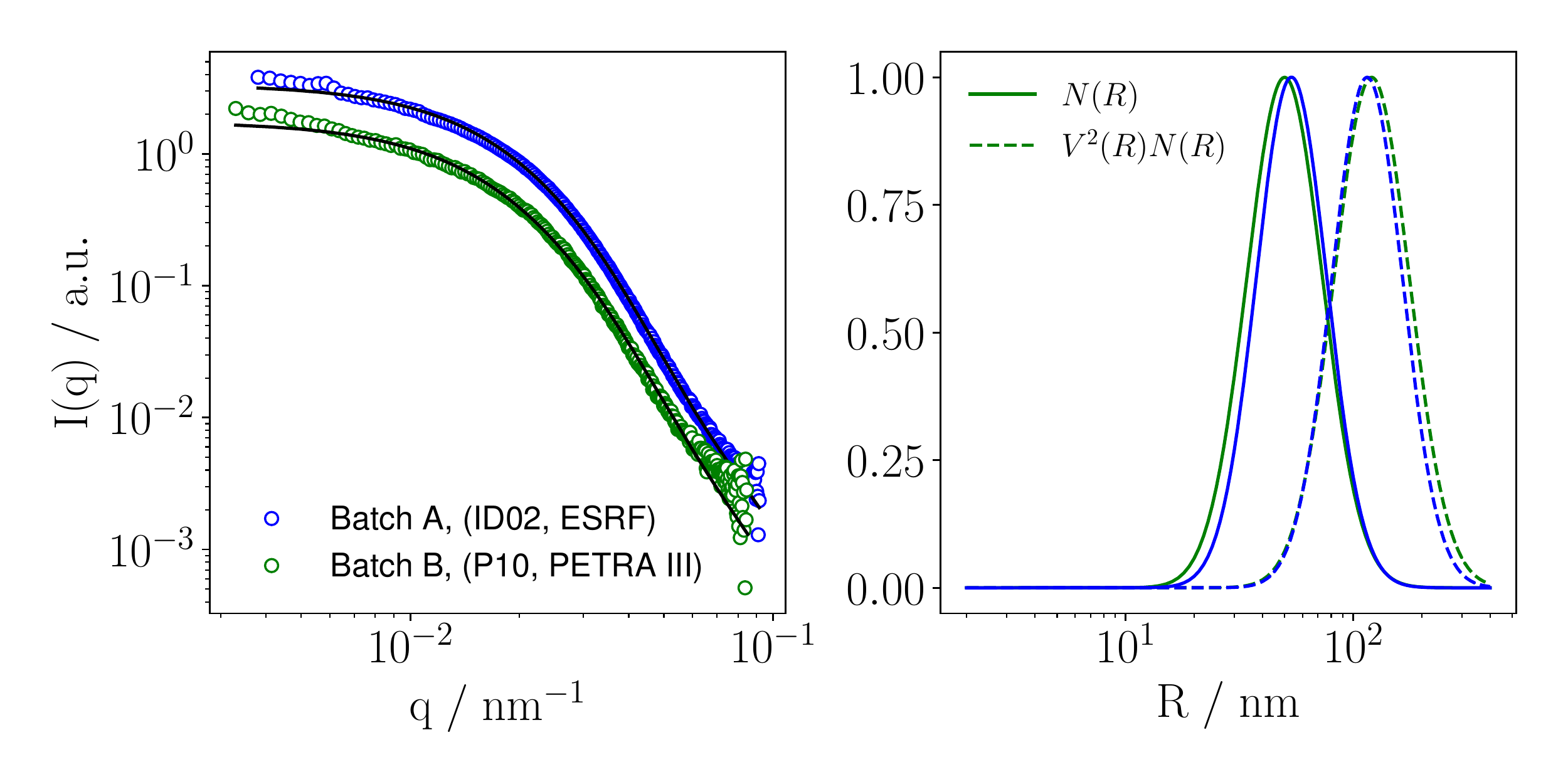}
  \caption{(left) SAXS intensities measured for two different batches of caseine micelles. (right) Size distribution of the casein micelles as obtained from fitting equation \ref{eq:poly} to the data in the left panel (solid lines). The dashed lines show the intensity weighted size distributions.}
  \label{fgr:saxs}
\end{figure}

The averaged SAXS intensity is displayed in figure \ref{fgr:saxs} (left). No structural changes were found as a function of exposure time and only slight differences were seen for two different batches which were measured at two different instruments, respectively. The SAXS data was modelled by a polydisperse sphere model
\begin{equation}
    I(q) \propto \int_0^{\infty} P(q,R) \ N(R) \ V^2(R) \ \mathrm{d}R, \label{eq:poly}
\end{equation}
with the form factor given by equation \ref{eq:form} and the distribution of sizes modelled as a logarithmic normal distribution
\begin{equation}
    N(R) = \frac{1}{\sqrt{2 \pi} R \ \sigma} \exp \left( - \frac{(\ln R - \mu)^2}{2 \sigma^2} \right).
\end{equation}
The resulting size distributions are displayed in figure \ref{fgr:saxs} (right). From these, mean radii can be obtained as $R_A = 65$ nm and $R_B = 62$ nm, which is close to values reported in comparable studies ($R \approx 70$ nm) \cite{Shukla2009}. Additionally, we retrieve the mean radii of the intensity weighted distribution function ($N_I(R) \propto V^2(R) \ N(R)$, dashed lines) as $R^{I}_A = 139$ nm and $R^{I}_B = 149$ nm.

\begin{figure}
  \includegraphics[width=0.95\textwidth]{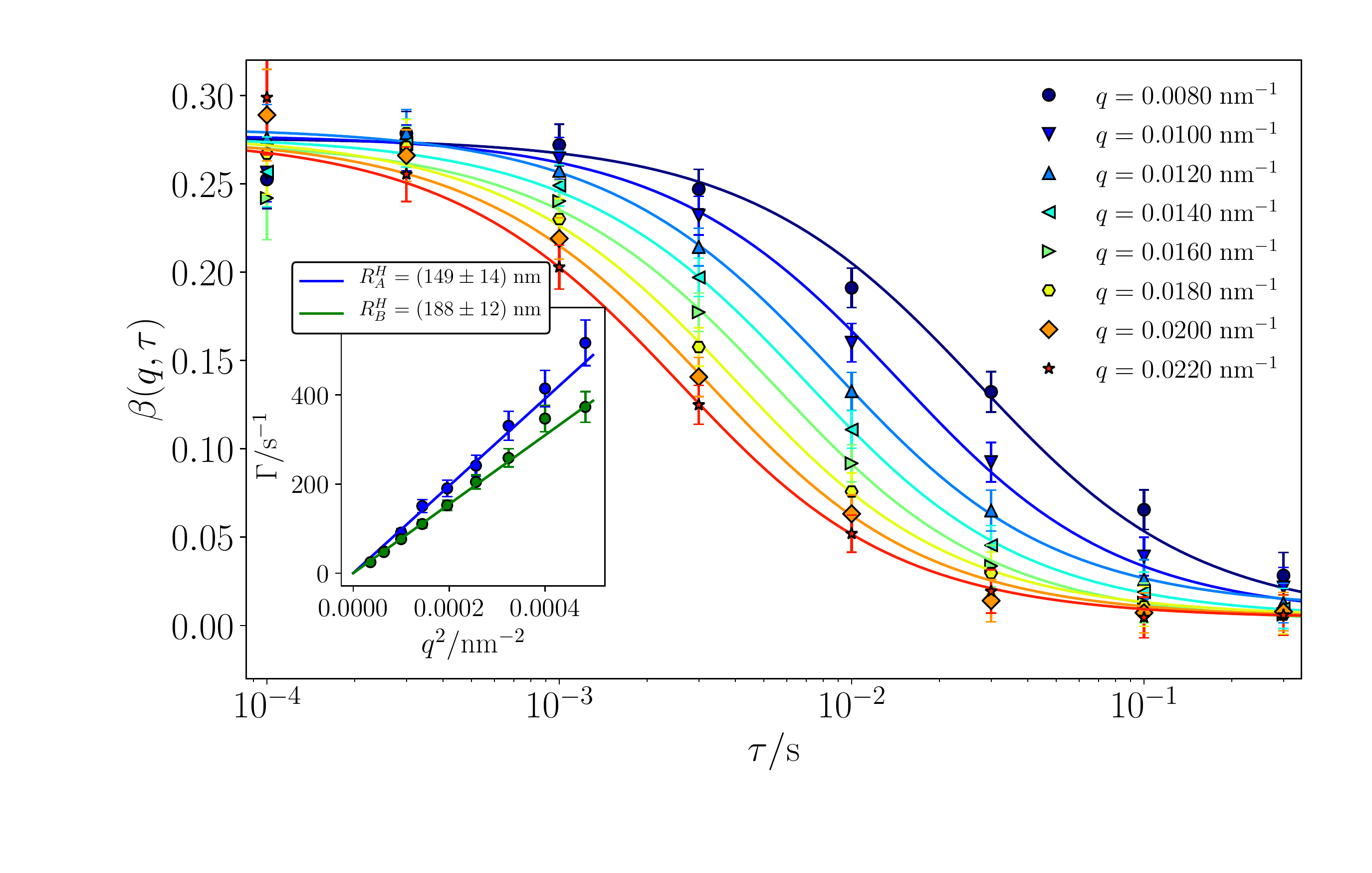}
  \caption{Speckle contrast $\beta(q,\tau)$ for a native casein solution, measured as a function of $q$ and exposure time $\tau$ at beamline ID02 (batch A). Solid lines present fits of equation \ref{eq:xsvs}, describing diffusive motion. The inset shows the retrieved relaxation rates $\Gamma(q)$, measured at two different beamlines (batch A and B, respectively). A linear dependence of $\Gamma$ with $q^2$ can be observed, characteristic for simple diffusive motion.}
  \label{fgr:cc}
\end{figure}

The measured speckle contrast $\beta(q,\tau)$ for the data set of batch A is displayed in Fig. \ref{fgr:cc}. We observe a clear dependence of $\beta$ on exposure time $\tau$ and on $q$ with high statistical accuracy, reflecting the dynamical behavior of the sample. For a quantitative evaluation, the data points are fitted following \cite{Bandy2005}
\begin{equation}
\beta(q,\tau) = \beta_0 \frac{ \exp(-2 \tau \Gamma(q))-1 + 2\tau \Gamma(q)}{2(\tau \Gamma(q))^2}, \label{eq:xsvs}
\end{equation}
which corresponds to an ISF of a system exhibiting diffusive motion described by an exponential decay
\begin{equation}
\left|g_{1}(q,t^{\prime})\right|  = \exp(- t^{\prime} \Gamma(q)).
\end{equation}
Here the relaxation rate $\Gamma(q)$ is given by $\Gamma(q)=  D_0 q^2$, so that the free diffusion coefficient $D_0$ can be obtained from the linear refinement as shown in the inset of Fig. \ref{fgr:cc}.\\
The hydrodynamic radius is given by
\begin{equation}
R_H = \frac{k_B T}{6 \pi \eta D_0},
\end{equation}
with $T$ and $k_B$ being the temperature and the Boltzmann constant, respectively. The viscosity $\eta = 1.5$ cP of skimmed milk was taken from literature  \cite{deKruif1999,Jeurnink1993}. Finally, we obtain $R^H_A = (149 \pm 14)$ nm (corresponding to the displayed correlation function) and $R^H_B = (188 \pm 12)$ nm. 

In order to compare the results from SAXS and XSVS, two influences on the obtained sizes have to be taken into account. First, casein micelles consist of a stabilizing brush layer ($\approx 11$ nm \cite{Shukla2009}), which due to its high hydration has only low contrast in SAXS but contributes to the apparent hydrodynamic radius. Therefore, we expect to measure a correspondingly larger hydrodynamic radius from XSVS as compared to the mean radius obtained by SAXS. Second, due to the large polydispersity of casein micelles, the measured hydrodynamic radius by XSVS is the mean radius of the intensity weighted size distribution and should in consequence be compared to $R^{I} \approx$ $139$ - $149$ nm. Therefore, we conclude that there is good agreement between both techniques.

\subsection{Dynamics upon acidification}
Manipulating the structure and stability of casein micelle solutions is well established and essential for applications like cheese or yogurt production \cite{deKruif1991,deKruif1992,Schurtenberger2001,pignon2004}. For cheese this is in general achieved by enzymatic cutting of the brushed $\kappa$-casein surface layer, denoted renneting. The hairy outer layer can also be decharged and ultimately collapsed by lowering the solution pH. This is for example achieved by the use of actio lactobacilli bacteria as in yogurt production.

\begin{figure}
  \includegraphics[width=0.95\textwidth]{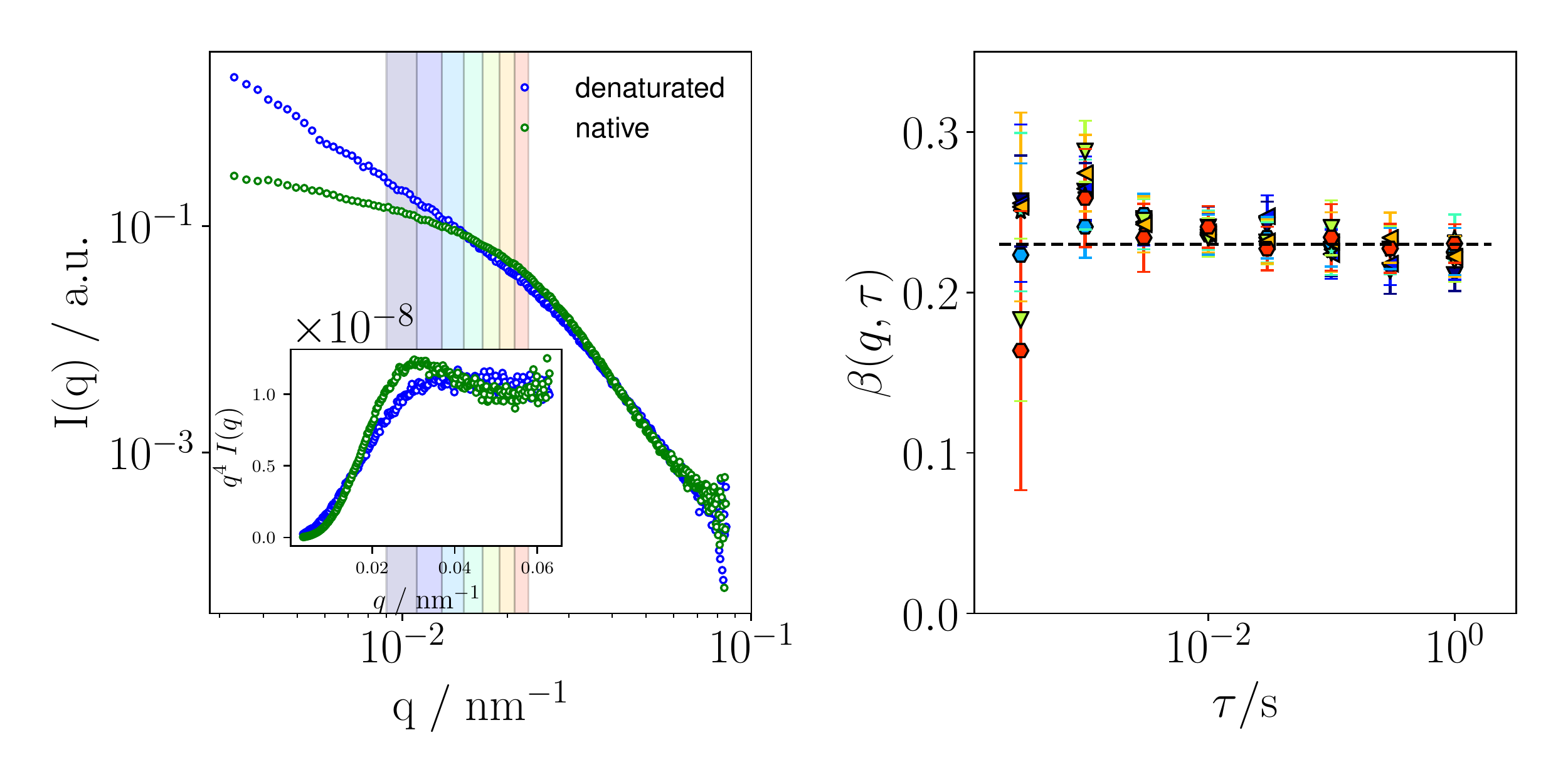}
  \caption{(left) SAXS intensities measured from a native (green) and an acid denatured (blue) skim milk sample, obtained at beamline P10. Color shaded regions depict the $q$-ranges for which the speckle contrast was analyzed. The inset displays the same data in a Porod representation. (right) Speckle contrast $\beta$ of an acidified skim milk sample as a function of $q$-bin and exposure time. The symbol colors follow the $q$-bins (left). The black dashed line depicts $\beta_0$ obtained from the fits to the native sample.}
  \label{fgr:hcl}
\end{figure}

We create such a denaturated state by changing the solution pH to 4.8 via careful titration with HCl and equilibrating the sample for 48 hours. The structural change witnessed by the SAXS curves are displayed in figure \ref{fgr:hcl} (left). Different influences of acidification on the structure and dynamics of casein micelles are reported in literature \cite{dalgleish1988,dalgleish1989,moitzi2011}. Due to the reduced pH, the charge on the repulsive stabilizing layer decreases. This lowers the colloidal stability and leads to aggregation of the sample, which can be observed by an upturn at low $q$ values in the SAXS intensity. Additionally, the decreased pH is known to dissociate the calcium phosphate in the micelles and therefore destabilizing the internal structure. This leads to a reduced globular compactness as evidenced by the reduced peak in the Porod plot (inset, Fig. \ref{fgr:hcl} left). However, a certain degree of structural integrity of the casein micelles remains, indicated by the preserved $q^{-4}$ decay of the scattering (corresponding to a plateau in the Porod plot at high $q$). As a consequence, certain deviations from purely colloidal gels are expected and observed for acidified casein gels \cite{horne2003}.

From an experimental point of view, especially in the broader context of aggregate  formation, phase transitions, solidification and gelation, beam damage effects can quickly accumulate due to reduced diffusion and exchange. As a result, such investigations are especially prone to radiation damage effects in sequential measurement schemes. Here, we demonstrate that even completely  arrested samples can be measured with XSVS, without signatures of radiation damage effects.

The corresponding $\beta(q,\tau)$ values are displayed in figure \ref{fgr:hcl} (right). Due to the long equilibration time, we expect the sample to have reached a kinetically arrested state and indeed we find even for the longest exposure times $\beta(q,\tau) \approx \beta_0$. We consider this an important demonstration of the measurement scheme being completely free of beam damage effects, since any beam induced influence to the sample would blur the speckle pattern with increasing exposure time. Potentially, this would allow to measure out-of-equilibrium conditions in-situ, following simultaneously the structure and dynamics after the pH induced transition.

\subsection{Length scale dependent dynamics}

An important feature of X-ray based, multi-speckle scattering techniques is the possibility of following structural and $q$-dependent dynamical information simultaneously. In concentrated solutions exhibiting strong interparticle interactions, the molecular displacements deviate from a Gaussian distribution, yielding a $q$-dependent diffusion coefficient $D(q)$ dominated by both static correlations and hydrodynamic protein-protein interactions mediated by the solvent \cite{DOSTER20071360}. Measurements of $D(q)$ allow distinguishing the different types of interactions by the short- and long-time diffusion coefficients at varying length scales.


\begin{figure}
  \includegraphics[width=0.95\textwidth]{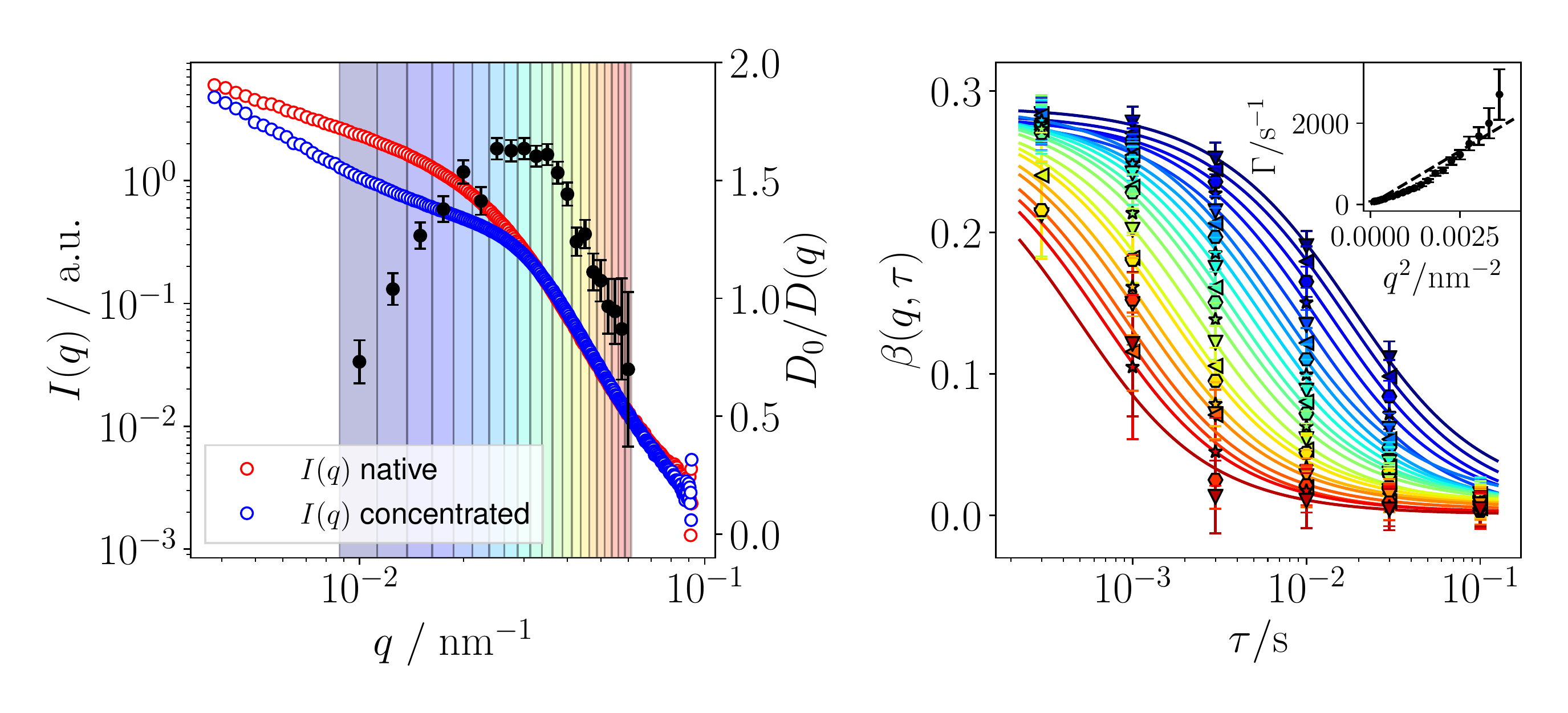}
  \caption{XSVS from a concentrated skim milk sample, obtained at beamline ID02. (left) SAXS intensities measured from the native (red) and the concentrated sample (blue). Black points: $D_0/D(q)$ with $D(q)$ denoting the $q$-dependent diffusion coefficient. Color shaded regions depict the $q$-ranges for which the speckle contrast was analyzed. (right) Speckle contrast $\beta$ of concentrated skim milk sample as a function of $q$-bin and exposure time. The symbol colors follow the $q$-bins (left). The solid lines display fits of eq. 8 to the data. The inset shows the relaxation rates $\Gamma(q^2)$, with a linear fit depicted as black dashed line.}
  \label{fgr:conc}
\end{figure}

We demonstrate the capability of determining $D(q)$  with the presented measurement scheme by using a concentrated skim milk solution, which was produced by dissolving three times the amount of skim milk powder as compared to the native sample. The measured SAXS curve is plotted in figure \ref{fgr:conc} (left), together with the $I(q)$ of the native sample. The formation of a shoulder at around $q = 3 \ 10^{-2} \ \mathrm{nm}^{-1}$ can be observed, stemming from the increasing direct particle interactions and thus an evolving static structure factor is present in the scattering intensity.

The corresponding $\beta(q,\tau)$ values are displayed in figure \ref{fgr:conc} (right). In order to expand the measured decay rates over a wider $q$-range, we additionally include $q$-bins where the lowest exposure times did not provide enough intensity for a reliable contrast determination or where the longest exposure times are not in the sparse scattering regime anymore. Therefore, we set a threshold of $3 \ 10^{-3} < \langle k \rangle < 10$, and exclude data points outside of this range from the plot and the fit.

The data points are fitted following equation \ref{eq:xsvs}, and the obtained relaxation rate $\Gamma$ is plotted as a function of $q^2$ in the inset of figure \ref{fgr:conc} (right). The relaxation rate of the concentrated solution displays clear deviation from the Brownian behavior ($\Gamma \propto q^2$, black dashed line). This deviation is additionally depicted in figure \ref{fgr:conc} (left) as $D_0 / D(q)$ (black data points).  $D_0 / D(q)$ primarily reflects the $q$-dependent static structure factor $S(q)$ following $D(q) = D_0H(q)/S(q)$, where $D_0$ is the Brownian diffusion coefficient of the diluted solution, modified by the hydrodynamic function $H(q)$, which captures the effect of hydrodynamic interactions \cite{Banchio2008,Bucciarellie1601432}.
The peak in $D_0 / D(q)$ thus resembles the peak in $S(q)/H(q)$ and indeed the postions of the shoulder in the SAXS curve and the peak in $D_0 / D(q)$ coincide at $q = 3 \ 10^{-2} \ \mathrm{nm}^{-1}$.

\section{Extension of Bio-XSVS to smaller length scales}
Having demonstrated that the dynamics of radiation sensitive samples can be measured with low dose XSVS, we now estimate the feasibility of our approach to cover an extended $q$-range corresponding to relevant length scales from several hundreds of nanometer down to a few nanometer. The crucial quantity for such experiments is how much dose $\tilde{D}$ is required on the sample in order to collect a sufficiently high intensity $\langle k \rangle$ per pixel and acquisition, while at the same time conserving the speckle contrast $\beta_0$. The quantity $\tilde{D}/\langle k \rangle$ can therefore be understood as a relative cost merit, which should be minimized in order to optimize the experimental conditions.

\begin{figure}
  \includegraphics[width=0.75\textwidth]{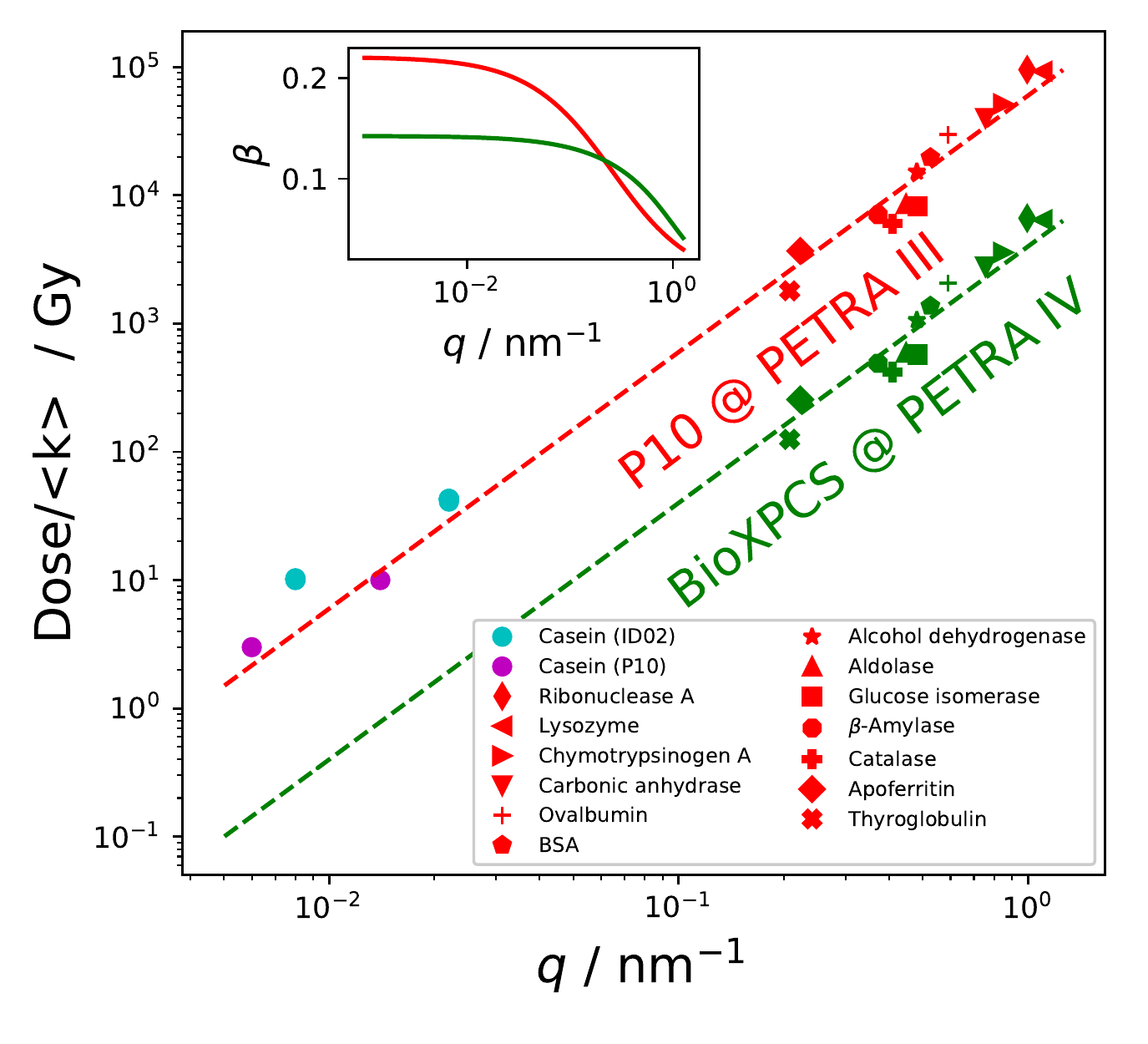}
  \caption{Comparison of different experimental setups in terms of deposited dose on the sample per mean pixel intensity $\langle k \rangle$. Measured values for Casein at beamline P10 (PETRA III) and ID02 (ESRF, prior to the ESRF-EBS upgrade) (cyan and magenta data points). Red data points are estimated values at $q/4$ where $q=2 \pi / R_G$  from measured absolute scattering intensities of different proteins \cite{Mylonas2007}, calculated with the same experimental parameter as used in this experiment at beamline P10. Green data points correspond to estimations using a proposed beamline design for biological coherent scattering applications (sample-detector-distance $L = 25$m, photon energy $E=16$ keV, beam spot size $a=50 \mu$m, Si-311 monochromator) as well as the improved source properties expected for the upgraded PETRA IV storage ring.}
  \label{fgr:dia}
\end{figure}

In a first approach, we assume that with the same USAXS setups (e.g. beamline P10, PETRA III: sample-detector-distance $L = 20$m, photon energy $E=8$ keV, beam spot size $a=60 \ \mu$m, Si-111 monochromator, EIGER detector) we can cover all required $q$-values. This approach can only serve as a lower limit estimation, as it already neglects geometrical limitations such as finite detector sizes or the restriction to a forward scattering geometry\footnote{In practice the required $q$-values could be reached also by a reduction of the sample-detector-distance, which would however require in consequence a smaller X-ray focus and therefore leading to a larger $\tilde{D}/\langle k \rangle$}. Still, we can estimate the theoretical $\tilde{D}/\langle k \rangle$ ratio at $q$-values corresponding to intermolecular distances for each protein ($q = 2 \pi / (4 \cdot R_G)$), using reported values for the radius of gyration $R_G$ and the absolute scattering intensity of each protein \cite{Mylonas2007} (red data points). Details of the calculation were reported in \cite{Moeller2019a}. All the data points follow approximately a $q^{2}$ behavior, provided as a guide to the eye (red dashed line in fig. \ref{fgr:dia}). Additionally, we include the measured $\tilde{D}/\langle k \rangle$ values of the previously discussed casein micelles solutions (for the largest and smallest $q$-values measured at both beamlines respectively (cyan and magenta data points in Fig. \ref{fgr:dia})).

Considering a maximum tolerable Dose of $\tilde{D} = 100$ Gy and a minimum scattering intensity of $\langle k \rangle = 10^{-2}$, we estimate the criterion of $\tilde{D}/ \langle k \rangle  \leq 10^4$ Gy for successful Bio-XSVS experiments. It can be seen that many of the displayed proteins exceed this value. Additionally, the speckle contrast at these larger $q$-values is also considerably reduced, as displayed by the red solid line in the inset of Fig. \ref{fgr:dia}. Therefore, we conclude that a simple extension of the experimental scheme to smaller length scales is not possible without revisiting the experimental setup.

The new generation of diffraction limited storage rings (DLSRs) such as the recently inaugurated ESRF-EBS \cite{Raimondi2016} or the planned PETRA IV facility \cite{Einfeld2014,Weckert2015,Schroer2018} provide X-ray beams with a higher coherent fraction, larger coherence length and higher coherent flux, especially at higher photon energies. As the presented measurements were performed before the upgrade of the ESRF storage ring, we expect that already now measurements on casein micelles can be expanded considerably to shorter time scales at the ID02 beamline \cite{Narayan2018}, since due to the increased coherent flux ($> 10^{12} ph/s$) a speckle pattern with the same intensity can be obtained within a shorter acquisition time. The investigation of smaller, monomeric proteins would however still be very challenging, as the ratio of $\tilde{D}/\langle k \rangle$ is independent of the coherent flux and the experimental geometry is restricted to forward scattering at ID02.

The improved coherent properties of X-ray radiation provided by new DLSR sources also open up new possibilities for designing low-dose coherent scattering experiments, with the aim of spreading the absorbed X-ray energy over a larger sample volume. In a nutshell, the larger transversal coherence length allows to increase the X-ray spot size on the sample while conserving the speckle contrast. Due to the higher incident coherent flux, the longitudinal coherence length can be increased by using a Si-311 monochromator while keeping a sufficient coherent flux on the sample. This, in combination with a higher photon energy, allows to increase the sample thickness. A more detailed discussion on the optimization of the scattering setup can be found in \cite{Moeller2019a}.

The envisioned increase in sampled volume also requires a high resolution scattering geometry to resolve the consequently smaller speckle on the detector and at the same time cover the targeted $q$-range. For this purpose, we propose a new beamline design optimized for biological coherent scattering applications, featuring a 25 m long, rotatable (horizontal 2$\Theta$) detector arm. We repeat the previous calculations with this proposed setup (sample-detector-distance $L = 25$m, photon energy $E=16$ keV, beam spot size $a=50 \ \mu$m, Si-311 monochromator) and using the predicted source properties of PETRA IV \cite{Schroer2018}. The corresponding data points are plotted in green in Fig. \ref{fgr:dia}. One can observe that a decrease of the deposited dose per $\langle k \rangle$ by about a factor 15 can be achieved, while even increasing the speckle contrast at larger $q$-values (inset). With this, biological samples can be studied on the full $q$-range of relevance, keeping the deposited dose per scattered photon below the threshold of $\tilde{D}/ \langle k \rangle  \leq 10^4$ Gy for all considered cases.\\

\section{Conclusion}

We demonstrate that dynamical information on biological, protein based samples in solution can be obtained from low dose XSVS measurements, with deposited doses per acquisition being well below critical dose limits. It was shown that simple, Brownian diffusion can be characterized for a native suspension of casein micelles, and the retrieved dynamical information matches well with expectations based on structural information that was obtained simultaneously by SAXS. In the case of a static soft matter system, no dynamics induced by radiation damage was observed, confirming the possibilities of XSVS for radiation sensitive systems. Additionally, more complex solution environments can be studied, such that for instance diffusive dynamics characterized by $D(q)$ in a crowded solution can be accessed experimentally.

We additionally show that XSVS is an especially well suited technique in combination with new generation X-ray sources. Maybe counter-intuitively, the increased coherent flux in combination with an increased coherence length, increased photon energy as well as dedicated instrumentation can be used to reduce the dose on the sample per scattered photon $\tilde{D}/\langle k \rangle$ and enable the investigation of small, monomeric proteins. It should be emphasized that the increase in coherent flux by novel X-ray sources can be fully utilized by this technique, also for the dynamics investigations of biological, radiation sensitive samples. The fastest accessible time scales are not limited by the acquisition rate of the detector, but only by the strength of the signal and the ability to determine the speckle contrast of low intensity speckle patterns with high statistical accuracy. Instruments for coherent scattering using faster detectors \cite{Zhang2018,Zinn2018,Jo2021} and high repetition rate sources such as European XFEL \cite{Lehmkuehler2020, madsen2021} will make it possible to obtain large sets of data in shorter time, in order to gather the required statistics and standardize the presented measurement scheme. In this sense, also the development of multi mega pixel detectors with higher sensitivity for single photon detection will push the detection limit for accurate speckle contrast determination towards lower count rates, enabling faster timescales to be probed and reducing the dose on the sample even further. Therefore, XSVS in combination with new coherent x-ray sources and single photon sensitive detectors will make the dynamical investigation of smaller, faster, and more radiation sensitive biological samples feasible.





\bibliographystyle{unsrt}
\bibliography{iucr}

\begin{thebibliography}{10}

\bibitem{Vodnala2018}
Preeti Vodnala, Nuwan Karunaratne, Laurence Lurio, George~M. Thurston, Michael
  Vega, Elizabeth Gaillard, Suresh Narayanan, Alec Sandy, Qingteng Zhang,
  Eric~M. Dufresne, Giuseppe Foffi, Pawel Grybos, Piotr Kmon, Piotr Maj, and
  Robert Szczygiel.
\newblock Hard-sphere-like dynamics in highly concentrated alpha-crystallin
  suspensions.
\newblock {\em Phys. Rev. E}, 97:020601, Feb 2018.

\bibitem{Moeller2019a}
Johannes M{\"{o}}ller, Michael Sprung, Anders Madsen, and Christian Gutt.
\newblock {X-ray photon correlation spectroscopy of protein dynamics at nearly
  diffraction-limited storage rings}.
\newblock {\em IUCrJ}, 6(5):794--803, Sep 2019.

\bibitem{Begam2021}
Nafisa Begam, Anastasia Ragulskaya, Anita Girelli, Hendrik Rahmann,
  Sivasurender Chandran, Fabian Westermeier, Mario Reiser, Michael Sprung,
  Fajun Zhang, Christian Gutt, and Frank Schreiber.
\newblock Kinetics of network formation and heterogeneous dynamics of an egg
  white gel revealed by coherent x-ray scattering.
\newblock {\em Phys. Rev. Lett.}, 126:098001, Mar 2021.

\bibitem{girelli_microscopic_2021}
Anita Girelli, Hendrik Rahmann, Nafisa Begam, Anastasia Ragulskaya, Mario
  Reiser, Sivasurender Chandran, Fabian Westermeier, Michael Sprung, Fajun
  Zhang, Christian Gutt, and Frank Schreiber.
\newblock Microscopic {Dynamics} of {Liquid}-{Liquid} {Phase} {Separation} and
  {Domain} {Coarsening} in a {Protein} {Solution} {Revealed} by {X}-{Ray}
  {Photon} {Correlation} {Spectroscopy}.
\newblock {\em Physical Review Letters}, 126(13):138004, April 2021.

\bibitem{Einfeld2014}
Dieter Einfeld, Mark Plesko, and Joachim Schaper.
\newblock {First multi-bend achromat lattice consideration}.
\newblock {\em Journal of Synchrotron Radiation}, 21(5):856--861, Sep 2014.

\bibitem{Weckert2015}
Edgar Weckert.
\newblock {The potential of future light sources to explore the structure and
  function of matter}.
\newblock {\em IUCrJ}, 2(2):230--245, Mar 2015.

\bibitem{Raimondi2016}
Pantaleo Raimondi.
\newblock Esrf ebs accelerator upgrade.
\newblock In {\em Proceedings of IPAC 2016}. 2016.

\bibitem{Tschentscher2017}
Thomas Tschentscher, Christian Bressler, Jan Gr{\"u}nert, Anders Madsen,
  Adrian~P Mancuso, Michael Meyer, Andreas Scherz, Harald Sinn, and Ulf
  Zastrau.
\newblock Photon beam transport and scientific instruments at the european
  xfel.
\newblock {\em Applied Sciences}, 7(6):592, 2017.

\bibitem{Schroer2018}
Christian~G. Schroer, Ilya Agapov, Werner Brefeld, Reinhard Brinkmann,
  Yong-Chul Chae, Hung-Chun Chao, Mikael Eriksson, Joachim Keil, Xavier
  Nuel~Gavald{\`a}, Ralf R{\"o}hlsberger, and et~al.
\newblock Petra iv: the ultralow-emittance source project at desy.
\newblock {\em Journal of Synchrotron Radiation}, 25(5):1277--1290, Aug 2018.

\bibitem{Decking2020}
W.~Decking, S.~Abeghyan, P.~Abramian, A.~Abramsky, A.~Aguirre, C.~Albrecht,
  P.~Alou, M.~Altarelli, P.~Altmann, K.~Amyan, V.~Anashin, E.~Apostolov,
  K.~Appel, D.~Auguste, V.~Ayvazyan, S.~Baark, F.~Babies, N.~Baboi, P.~Bak,
  V.~Balandin, R.~Baldinger, B.~Baranasic, S.~Barbanotti, O.~Belikov,
  V.~Belokurov, L.~Belova, V.~Belyakov, S.~Berry, M.~Bertucci, B.~Beutner, Anke
  Block, M.~Blöcher, T.~Böckmann, C.~Bohm, M.~Böhnert, V.~Bondar,
  E.~Bondarchuk, M.~Bonezzi, P.~Borowiec, C.~Bösch, U.~Bösenberg, A.~Bosotti,
  R.~Böspflug, M.~Bousonville, E.~Boyd, Y.~Bozhko, A.~Brand, J.~Branlard,
  S.~Briechle, F.~Brinker, S.~Brinker, R.~Brinkmann, S.~Brockhauser, O.~Brovko,
  H.~Brück, A.~Brüdgam, L.~Butkowski, T.~Büttner, J.~Calero,
  E.~Castro-Carballo, G.~Cattalanotto, J.~Charrier, J.~Chen, A.~Cherepenko,
  V.~Cheskidov, M.~Chiodini, A.~Chong, S.~Choroba, M.~Chorowski, D.~Churanov,
  W.~Cichalewski, M.~Clausen, W.~Clement, C.~Cloué, J.~A. Cobos, N.~Coppola,
  S.~Cunis, K.~Czuba, M.~Czwalinna, B.~D’Almagne, J.~Dammann, H.~Danared,
  A.~Delfs, T.~Delfs, F.~Dietrich, T.~Dietrich, M.~Dohlus, M.~Dommach,
  A.~Donat, X.~Dong, N.~Doynikov, Michael Dressel, M.~Duda, P.~Duda,
  H.~Eckoldt, W.~Ehsan, J.~Eidam, F.~Eints, C.~Engling, U.~Englisch, Alexey
  Ermakov, Kurt Escherich, J.~Eschke, E.~Saldin, M.~Faesing, A.~Fallou,
  M.~Felber, M.~Fenner, B.~Fernandes, J.~M. Fernández, S.~Feuker,
  K.~Filippakopoulos, K.~Floettmann, V.~Fogel, M.~Fontaine, A.~Francés,
  I.~Freijo Martin, W.~Freund, T.~Freyermuth, M.~Friedland, L.~Fröhlich,
  M.~Fusetti, J.~Fydrych, A.~Gallas, O.~García, L.~Garcia-Tabares, G.~Geloni,
  N.~Gerasimova, C.~Gerth, P.~Geßler, V.~Gharibyan, M.~Gloor,
  J.~Glowinkowski, A.~Goessel, Z.~Golebiewski, N.~Golubeva, W.~Grabowski,
  W.~Graeff, A.~Grebentsov, M.~Grecki, T.~Grevsmuehl, M.~Gross,
  U.~Grosse-Wortmann, Jan Gruenert, Soeren Grunewald, P.~Grzegory, G.~Feng,
  H.~Guler, G.~Gusev, J.~L. Gutierrez, L.~Hagge, M.~Hamberg, R.~Hanneken,
  E.~Harms, I.~Hartl, A.~Hauberg, S.~Hauf, J.~Hauschildt, Jens Hauser,
  J.~Havlicek, A.~Hedqvist, N.~Heidbrook, F.~Hellberg, D.~Henning, O.~Hensler,
  Tim Hermann, A.~Hidvégi, M.~Hierholzer, H.~Hintz, Frank Hoffmann, Markus
  Hoffmann, Matthias Hoffmann, Y.~Holler, Markus Hüning, A.~Ignatenko, Markus
  Ilchen, A.~Iluk, J.~Iversen, M.~Izquierdo, Lutz Schmidt~genannt Jachmann,
  N.~Jardon, U.~Jastrow, K.~Jensch, Jens-Peter Jensen, M.~Jeżabek, M.~Jidda,
  H.~Jin, N.~Johansson, R.~Jonas, W.~Kaabi, D.~Kaefer, R.~Kammering,
  H.~Kapitza, S.~Karabekyan, S.~Karstensen, K.~Kasprzak, V.~Katalev, D.~Keese,
  B.~Keil, M.~Kholopov, M.~Killenberger, B.~Kitaev, Y.~Klimchenko, R.~Klos,
  L.~Knebel, Andreas Koch, Michael Koepke, S.~Köhler, Winfried Koehler,
  N.~Kohlstrunk, Z.~Konopkova, A.~Konstantinov, W.~Kook, W.~Koprek, M.~Körfer,
  O.~Korth, A.~Kosarev, K.~Kosiński, D.~Kostin, Y.~Kot, Andrzej Kotarba,
  T.~Kozak, V.~Kozak, R.~Kramert, M.~Krasilnikov, A.~Krasnov, Bernward Krause,
  L.~Kravchuk, O.~Krebs, R.~Kretschmer, J.~Kreutzkamp, O.~Kröplin, K.~Krzysik,
  G.~Kube, H.~Kuehn, N.~Kujala, V.~Kulikov, V.~Kuzminych, D.~La~Civita,
  M.~Lacroix, T.~Lamb, A.~Lancetov, M.~Larsson, D.~Le~Pinvidic, S.~Lederer,
  Timmy-Jan Lensch, Dennis Lenz, A.~Leuschner, F.~Levenhagen, Yuhui Li,
  J.~Liebing, L.~Lilje, T.~Limberg, D.~Lipka, B.~List, J.~Liu, Shan Liu,
  B.~Lorbeer, J.~Lorkiewicz, H.~H. Lu, F.~Ludwig, K.~Machau, W.~Maciocha,
  C.~Madec, C.~Magueur, C.~Maiano, I.~Maksimova, K.~Malcher, T.~Maltezopoulos,
  E.~Mamoshkina, B.~Manschwetus, F.~Marcellini, G.~Marinkovic, T.~Martinez,
  H.~Martirosyan, W.~Maschmann, Mikhail Maslov, A.~Matheisen, U.~Mavric, Joshua
  Meissner, K.~Meissner, M.~Messerschmidt, N.~Meyners, G.~Michalski,
  P.~Michelato, N.~Mildner, M.~Moe, F.~Moglia, C.~Mohr, Stefan Mohr,
  W.~Möller, M.~Mommerz, L.~Monaco, C.~Montiel, M.~Moretti, I.~Morozov, Petr
  Morozov, Dieter Mross, Joachim Mueller, Carsten Mueller, Jost Müller, Kurt
  Mueller, J.~Munilla, A.~Münnich, V.~Muratov, O.~Napoly, B.~Näser,
  N.~Nefedov, Reinhard Neumann, Rudolf Neumann, N.~Ngada, D.~Noelle, F.~Obier,
  I.~Okunev, J.~A. Oliver, M.~Omet, Anne Oppelt, A.~Ottmar, M.~Oublaid,
  C.~Pagani, R.~Paparella, V.~Paramonov, C.~Peitzmann, J.~Penning, A.~Perus,
  Falko Peters, Britta Petersen, Alexander Petrov, I.~Petrov, S.~Pfeiffer,
  J.~Pflüger, S.~Philipp, Y.~Pienaud, P.~Pierini, S.~Pivovarov, M.~Planas,
  E.~Pławski, Mario Pohl, J.~Polinski, V.~Popov, S.~Prat, J.~Prenting, Gunnar
  Priebe, H.~Pryschelski, K.~Przygoda, E.~Pyata, B.~Racky, A.~Rathjen,
  W.~Ratuschni, S.~Regnaud-Campderros, K.~Rehlich, D.~Reschke, C.~Robson,
  J.~Roever, M.~Roggli, J.~Rothenburg, E.~Rusiński, R.~Rybaniec, H.~Sahling,
  M.~Salmani, L.~Samoylova, D.~Sanzone, F.~Saretzki, O.~Sawlanski,
  J.~Schaffran, H.~Schlarb, M.~Schlösser, V.~Schlott, Christian Schmidt,
  F.~Schmidt-Foehre, Michael Schmitz, M.~Schmökel, T.~Schnautz,
  E.~Schneidmiller, Matthias Scholz, B.~Schöneburg, J.~Schultze, C.~Schulz,
  A.~Schwarz, J.~Sekutowicz, D.~Sellmann, E.~Semenov, S.~Serkez, D.~Sertore,
  N.~Shehzad, P.~Shemarykin, L.~Shi, M.~Sienkiewicz, D.~Sikora, M.~Sikorski,
  A.~Silenzi, C.~Simon, W.~Singer, X.~Singer, H.~Sinn, K.~Sinram,
  N.~Skvorodnev, P.~Smirnow, Thorsten Sommer, Andrey Sorokin, M.~Stadler,
  M.~Steckel, Bernd Steffen, N.~Steinhau-Kühl, F.~Stephan, M.~Stodulski, Maja
  Stolper, A.~Sulimov, R.~Susen, J.~Świerblewski, C.~Sydlo, E.~Syresin,
  V.~Sytchev, J.~Szuba, N.~Tesch, J.~Thie, A.~Thiebault, K.~Tiedtke,
  D.~Tischhauser, J.~Tolkiehn, S.~Tomin, F.~Tonisch, F.~Toral, I.~Torbin,
  A.~Trapp, D.~Treyer, G.~Trowitzsch, T.~Trublet, T.~Tschentscher, F.~Ullrich,
  M.~Vannoni, P.~Varela, G.~Varghese, Grygorii Vashchenko, M.~Vasic,
  C.~Vazquez-Velez, A.~Verguet, S.~Vilcins-Czvitkovits, R.~Villanueva,
  B.~Visentin, M.~Viti, E.~Vogel, E.~Volobuev, Richard Wagner, N.~Walker,
  T.~Wamsat, H.~Weddig, G.~Weichert, H.~Weise, R.~Wenndorf, M.~Werner,
  R.~Wichmann, C.~Wiebers, M.~Wiencek, T.~Wilksen, I.~Will, L.~Winkelmann,
  M.~Winkowski, Kay Wittenburg, A.~Witzig, P.~Wlk, T.~Wohlenberg,
  M.~Wojciechowski, F.~Wolff-Fabris, G.~Wrochna, K.~Wrona, M.~Yakopov, Bin
  Yang, Fan Yang, M.~Yurkov, I.~Zagorodnov, P.~Zalden, A.~Zavadtsev,
  D.~Zavadtsev, A.~Zhirnov, A.~Zhukov, V.~Ziemann, A.~Zolotov, N.~Zolotukhina,
  F.~Zummack, D.~Zybin, and Antonio~de Zubiaurre-Wagner.
\newblock {A} {MH}z-{R}epetition-{R}ate {H}ard {X}-{R}ay {F}ree-{E}lectron
  {L}aser {D}riven by a {S}uperconducting {L}inear {A}ccelerator.
\newblock {\em Nature photonics}, 14(6):391 -- 397, 2020.

\bibitem{Stradner2020}
Anna Stradner and Peter Schurtenberger.
\newblock Potential and limits of a colloid approach to protein solutions.
\newblock {\em Soft Matter}, 16:307--323, 2020.

\bibitem{grimaldo_2019}
Marco Grimaldo, Felix Roosen-Runge, Fajun Zhang, Frank Schreiber, and Tilo
  Seydel.
\newblock Dynamics of proteins in solution.
\newblock {\em Quarterly Reviews of Biophysics}, 52:e7, 2019.

\bibitem{Ruta2017}
Beatrice Ruta, F~Zontone, Y~Chushkin, G~Baldi, G~Pintori, G~Monaco, Benoit
  Ruffle, and Walter Kob.
\newblock Hard x-rays as pump and probe of atomic motion in oxide glasses.
\newblock {\em Scientific reports}, 7(1):3962, 2017.

\bibitem{Pintori2019}
G.~Pintori, G.~Baldi, B.~Ruta, and G.~Monaco.
\newblock Relaxation dynamics induced in glasses by absorption of hard x-ray
  photons.
\newblock {\em Phys. Rev. B}, 99:224206, Jun 2019.

\bibitem{Holzweber2019}
Katharina Holzweber, Christoph Tietz, Tobias~Michael Fritz, Bogdan Sepiol, and
  Michael Leitner.
\newblock Beam-induced atomic motion in alkali borate glasses.
\newblock {\em Phys. Rev. B}, 100:214305, Dec 2019.

\bibitem{Fluerasu2008}
Andrei Fluerasu, Abdellatif Moussa{\"\i}d, P{\'{e}}ter Falus, Henri Gleyzolle,
  and Anders Madsen.
\newblock {X-ray photon correlation spectroscopy under flow}.
\newblock {\em Journal of Synchrotron Radiation}, 15(4):378--384, Jul 2008.

\bibitem{westermeier2012}
Fabian Westermeier, Birgit Fischer, Wojciech Roseker, Gerhard Gr{\"u}bel,
  Gerhard N{\"a}gele, and Marco Heinen.
\newblock Structure and short-time dynamics in concentrated suspensions of
  charged colloids.
\newblock {\em The Journal of chemical physics}, 137(11):114504, 2012.

\bibitem{Lurio2021}
Laurence~B. Lurio, George~M. Thurston, Qingteng Zhang, Suresh Narayanan, and
  Eric~M. Dufresne.
\newblock {Use of continuous sample translation to reduce radiation damage for
  XPCS studies of protein diffusion}.
\newblock {\em Journal of Synchrotron Radiation}, 28(2), Mar 2021.

\bibitem{Verwohlt2018}
Jan Verwohlt, Mario Reiser, Lisa Randolph, Aleksandar Matic, Luis~Aguilera
  Medina, Anders Madsen, Michael Sprung, Alexey Zozulya, and Christian Gutt.
\newblock Low dose x-ray speckle visibility spectroscopy reveals nanoscale
  dynamics in radiation sensitive ionic liquids.
\newblock {\em Phys. Rev. Lett.}, 120:168001, Apr 2018.

\bibitem{Jeffries2015}
{C.M.} Jeffries, {M.A.} Graewert, {D.I.} Svergun, and {C.E.} Blanchet.
\newblock Limiting radiation damage for high-brilliance biological solution
  scattering: practical experience at the {EMBL} p12 beamline {PETRAIII}.
\newblock 22(2):273--279, 2015.

\bibitem{Gutt2009}
C~Gutt, L-M Stadler, Agn{\`e}s Duri, T~Autenrieth, O~Leupold, Y~Chushkin, and
  G~Gr{\"u}bel.
\newblock Measuring temporal speckle correlations at ultrafast x-ray sources.
\newblock {\em Optics express}, 17(1):55--61, 2009.

\bibitem{Hruszkewycz2012}
SO~Hruszkewycz, M~Sutton, PH~Fuoss, B~Adams, S~Rosenkranz, KF~Ludwig~Jr,
  W~Roseker, D~Fritz, M~Cammarata, D~Zhu, et~al.
\newblock High contrast x-ray speckle from atomic-scale order in liquids and
  glasses.
\newblock {\em Physical review letters}, 109(18):185502, 2012.

\bibitem{Inoue2012}
Ichiro Inoue, Yuya Shinohara, Akira Watanabe, and Yoshiyuki Amemiya.
\newblock Effect of shot noise on x-ray speckle visibility spectroscopy.
\newblock {\em Optics express}, 20(24):26878--26887, 2012.

\bibitem{Li2014}
Luxi Li, Pawe{\l} Kwa{\'s}niewski, Davide Orsi, Lutz Wiegart, Luigi
  Cristofolini, Chiara Caronna, and Andrei Fluerasu.
\newblock Photon statistics and speckle visibility spectroscopy with partially
  coherent x-rays.
\newblock {\em Journal of synchrotron radiation}, 21(6):1288--1295, 2014.

\bibitem{Zinn2018}
T.~Zinn, A.~Homs, L.~Sharpnack, G~Tinti, E~Fr{\"{o}}jdh, P.-A. Douissard,
  M.~Kocsis, J.~M{\"{o}}ller, Y.~Chushkin, and T.~Narayanan.
\newblock {Ultra-small-angle X-ray photon correlation spectroscopy using the
  Eiger detector}.
\newblock {\em Journal of Synchrotron Radiation}, 25(6):1753--1759, Nov 2018.

\bibitem{Dinapoli2011}
Roberto Dinapoli, Anna Bergamaschi, Beat Henrich, Roland Horisberger, Ian
  Johnson, Aldo Mozzanica, Elmar Schmid, Bernd Schmitt, Akos Schreiber, Xintian
  Shi, et~al.
\newblock Eiger: Next generation single photon counting detector for x-ray
  applications.
\newblock {\em Nuclear Instruments and Methods in Physics Research Section A:
  Accelerators, Spectrometers, Detectors and Associated Equipment},
  650(1):79--83, 2011.

\bibitem{Radicci2012}
V~Radicci, A~Bergamaschi, R~Dinapoli, D~Greiffenberg, B~Henrich, I~Johnson,
  A~Mozzanica, B~Schmitt, and X~Shi.
\newblock Eiger a new single photon counting detector for x-ray applications:
  performance of the chip.
\newblock {\em Journal of Instrumentation}, 7(02):C02019, 2012.

\bibitem{Johnson2014}
I~Johnson, Anna Bergamaschi, Heiner Billich, S~Cartier, R~Dinapoli, Dominic
  Greiffenberg, Manuel Guizar-Sicairos, B~Henrich, J~Jungmann, Davide Mezza,
  et~al.
\newblock Eiger: a single-photon counting x-ray detector.
\newblock {\em Journal of Instrumentation}, 9(05):C05032, 2014.

\bibitem{Bandy2005}
Ranjini Bandyopadhyay, AS~Gittings, SS~Suh, PK~Dixon, and Douglas~J Durian.
\newblock Speckle-visibility spectroscopy: A tool to study time-varying
  dynamics.
\newblock {\em Review of scientific instruments}, 76(9):093110, 2005.

\bibitem{Goodman1985}
J.W. Goodman.
\newblock Statistical optics.
\newblock In {\em Statistical Optics}. Wiley, New York, 1985.

\bibitem{Moeller2019b}
Johannes M{\"{o}}ller, Mario Reiser, J{\"{o}}rg Hallmann, Ulrike Boesenberg,
  Alexey Zozulya, Hendrik Rahmann, Anna-Lena Becker, Fabian Westermeier, Thomas
  Zinn, Federico Zontone, Christian Gutt, and Anders Madsen.
\newblock {Implications of disturbed photon-counting statistics of Eiger
  detectors for X-ray speckle visibility experiments}.
\newblock {\em Journal of Synchrotron Radiation}, 26(5):1705--1715, Sep 2019.

\bibitem{horne2006}
David~S Horne.
\newblock Casein micelle structure: models and muddles.
\newblock {\em Current opinion in colloid \& interface science},
  11(2-3):148--153, 2006.

\bibitem{Shukla2009}
Anuj Shukla, Theyencheri Narayanan, and Dra{\v{z}}en Zanchi.
\newblock Structure of casein micelles and their complexation with tannins.
\newblock {\em Soft Matter}, 5(15):2884--2888, 2009.

\bibitem{deKruif2012}
Cornelis~G de~Kruif, Thom Huppertz, Volker~S Urban, and Andrei~V Petukhov.
\newblock Casein micelles and their internal structure.
\newblock {\em Advances in colloid and interface science}, 171:36--52, 2012.

\bibitem{Ingham2016}
Bridget Ingham, Alice Smialowska, Gad~Damar Erlangga, Lara Matia-Merino,
  NM~Kirby, Cheng Wang, Richard~G Haverkamp, and AJ~Carr.
\newblock Revisiting the interpretation of casein micelle saxs data.
\newblock {\em Soft Matter}, 12(33):6937--6953, 2016.

\bibitem{mezzenga2005}
Raffaele Mezzenga, Peter Schurtenberger, Adam Burbidge, and Martin Michel.
\newblock Understanding foods as soft materials.
\newblock {\em Nature materials}, 4(10):729--740, 2005.

\bibitem{urban1999}
Claus Urban and Peter Schurtenberger.
\newblock Application of a new light scattering technique to avoid the
  influence of dilution in light scattering experiments with milk.
\newblock {\em Physical Chemistry Chemical Physics}, 1(17):3911--3915, 1999.

\bibitem{dahbi2010}
Louisa Dahbi, M~Alexander, V{\'e}ronique Trappe, JKG Dhont, and Peter
  Schurtenberger.
\newblock Rheology and structural arrest of casein suspensions.
\newblock {\em Journal of Colloid and Interface Science}, 342(2):564--570,
  2010.

\bibitem{alexander2008}
Marcela Alexander, Ivo Piska, and Douglas~G Dalgleish.
\newblock Investigation of particle dynamics in gels involving casein micelles:
  a diffusing wave spectroscopy and rheology approach.
\newblock {\em Food hydrocolloids}, 22(6):1124--1134, 2008.

\bibitem{deKruif1999}
CG~de~Kruif.
\newblock Casein micelle interactions.
\newblock {\em International Dairy Journal}, 9(3-6):183--188, 1999.

\bibitem{Jeurnink1993}
Theo~JM Jeurnink and Kees~G De~Kruif.
\newblock Changes in milk on heating: viscosity measurements.
\newblock {\em Journal of Dairy Research}, 60(2):139--150, 1993.

\bibitem{deKruif1991}
Cees~G de~Kruif and Roland~P May.
\newblock $\kappa$-casein micelles: structure, interaction and gelling studied
  by small-angle neutron scattering.
\newblock {\em European journal of biochemistry}, 200(2):431--436, 1991.

\bibitem{deKruif1992}
CG~de~Kruif.
\newblock Casein micelles: diffusivity as a function of renneting time.
\newblock {\em Langmuir}, 8(12):2932--2937, 1992.

\bibitem{Schurtenberger2001}
Peter Schurtenberger, Anna Stradner, Sara Romer, Claus Urban, and Frank
  Scheffold.
\newblock Aggregation and gel formation in biopolymer solutions.
\newblock {\em CHIMIA International Journal for Chemistry}, 55(3):155--159,
  2001.

\bibitem{pignon2004}
Fr{\'e}d{\'e}ric Pignon, Gabor Belina, Theyencheri Narayanan, Xavier Paubel,
  Albert Magnin, and Genevi{\`e}ve G{\'e}san-Guiziou.
\newblock Structure and rheological behavior of casein micelle suspensions
  during ultrafiltration process.
\newblock {\em The Journal of chemical physics}, 121(16):8138--8146, 2004.

\bibitem{dalgleish1988}
Douglas~G Dalgleish and Andrew~JR Law.
\newblock ph-induced dissociation of bovine casein micelles. i. analysis of
  liberated caseins.
\newblock {\em Journal of Dairy Research}, 55(4):529--538, 1988.

\bibitem{dalgleish1989}
Douglas~G Dalgleish and Andrew~JR Law.
\newblock ph-induced dissociation of bovine casein micelles. ii. mineral
  solubilization and its relation to casein release.
\newblock {\em Journal of Dairy Research}, 56(5):727--735, 1989.

\bibitem{moitzi2011}
Christian Moitzi, Andreas Menzel, Peter Schurtenberger, and Anna Stradner.
\newblock The ph induced sol- gel transition in skim milk revisited. a detailed
  study using time-resolved light and x-ray scattering experiments.
\newblock {\em Langmuir}, 27(6):2195--2203, 2011.

\bibitem{horne2003}
David~S Horne.
\newblock Casein micelles as hard spheres: limitations of the model in
  acidified gel formation.
\newblock {\em Colloids and Surfaces A: Physicochemical and Engineering
  Aspects}, 213(2-3):255--263, 2003.

\bibitem{DOSTER20071360}
Wolfgang Doster and Stéphane Longeville.
\newblock Microscopic diffusion and hydrodynamic interactions of hemoglobin in
  red blood cells.
\newblock {\em Biophysical Journal}, 93(4):1360--1368, 2007.

\bibitem{Banchio2008}
Adolfo~J. Banchio and Gerhard Nägele.
\newblock Short-time transport properties in dense suspensions: From neutral to
  charge-stabilized colloidal spheres.
\newblock {\em The Journal of Chemical Physics}, 128(10):104903, 2008.

\bibitem{Bucciarellie1601432}
Saskia Bucciarelli, Jin~Suk Myung, Bela Farago, Shibananda Das, Gerard~A.
  Vliegenthart, Olaf Holderer, Roland~G. Winkler, Peter Schurtenberger, Gerhard
  Gompper, and Anna Stradner.
\newblock Dramatic influence of patchy attractions on short-time protein
  diffusion under crowded conditions.
\newblock {\em Science Advances}, 2(12), 2016.

\bibitem{Mylonas2007}
Efstratios Mylonas and Dmitri~I. Svergun.
\newblock {Accuracy of molecular mass determination of proteins in solution by
  small-angle X-ray scattering}.
\newblock {\em Journal of Applied Crystallography}, 40(s1):s245--s249, Apr
  2007.

\bibitem{Narayan2018}
Theyencheri Narayanan, Michael Sztucki, Pierre Van~Vaerenbergh, Joachim
  L{\'{e}}onardon, Jacques Gorini, Laurent Claustre, Franc Sever, John Morse,
  and Peter Boesecke.
\newblock {A multipurpose instrument for time-resolved ultra-small-angle and
  coherent X-ray scattering}.
\newblock {\em Journal of Applied Crystallography}, 51(6):1511--1524, Dec 2018.

\bibitem{Zhang2018}
Qingteng Zhang, Eric~M. Dufresne, Suresh Narayanan, Piotr Maj, Anna Koziol,
  Robert Szczygiel, Pawel Grybos, Mark Sutton, and Alec~R. Sandy.
\newblock {Sub-microsecond-resolved multi-speckle X-ray photon correlation
  spectroscopy with a pixel array detector}.
\newblock {\em Journal of Synchrotron Radiation}, 25(5):1408--1416, Sep 2018.

\bibitem{Jo2021}
Wonhyuk Jo, Fabian Westermeier, Rustam Rysov, Olaf Leupold, Florian Schulz,
  Steffen Tober, Verena Markmann, Michael Sprung, Allesandro Ricci, Torsten
  Laurus, et~al.
\newblock Nanosecond x-ray photon correlation spectroscopy using pulse time
  structure of a storage-ring source.
\newblock {\em IUCrJ}, 8(1), 2021.

\bibitem{Lehmkuehler2020}
Felix Lehmk{\"u}hler, Francesco Dallari, Avni Jain, Marcin Sikorski, Johannes
  M{\"o}ller, Lara Frenzel, Irina Lokteva, Grant Mills, Michael Walther, Harald
  Sinn, Florian Schulz, Michael Dartsch, Verena Markmann, Richard Bean, Yoonhee
  Kim, Patrik Vagovic, Anders Madsen, Adrian~P. Mancuso, and Gerhard
  Gr{\"u}bel.
\newblock Emergence of anomalous dynamics in soft matter probed at the european
  xfel.
\newblock {\em Proceedings of the National Academy of Sciences},
  117(39):24110--24116, 2020.

\bibitem{madsen2021}
A~Madsen, J~Hallmann, G~Ansaldi, T~Roth, W~Lu, C~Kim, U~Boesenberg, A~Zozulya,
  J~M{\"o}ller, R~Shayduk, et~al.
\newblock Materials imaging and dynamics (mid) instrument at the european x-ray
  free-electron laser facility.
\newblock {\em Journal of Synchrotron Radiation}, 28(2), 2021.

\end{thebibliography}






\end{document}